\documentclass[sigconf]{acmart}
\usepackage{array}
\AtBeginDocument{%
  }

\setcopyright{acmlicensed}
\copyrightyear{2026}
\acmYear{2026}
\acmDOI{XXXXXXX.XXXXXXX}
\acmConference[CHI EA '26]{the 2026 CHI Workshop on Human-AI Interaction Alignment: Designing, Evaluating, and Evolving Value-Centered AI For Reciprocal Human-AI Futures}{April 13--17, 2026}{Barcelona, Spain}

\acmISBN{978-1-4503-XXXX-X/2018/06}

\begin{document}

\title{Human Agency, Causality, and the Human Computer Interface in High-Stakes Artificial Intelligence}

\newcommand{\RKI}{Center for Artificial Intelligence in Public Health Research (ZKI-PH), Robert Koch Institute}
\newcommand{\FUB}{Department of Mathematics and Computer Science, Freie Universität Berlin}

\author{Georges Hattab}
\email{hattabg@rki.de; georges.hattab@fu-berlin.de}
\orcid{1234-5678-9012}
\affiliation{%
\institution{\RKI;}
\institution{\FUB}
  \city{Berlin}
  \country{Germany}
}


\begin{abstract}
Current discourse on Artificial Intelligence (AI) ethics, dominated by ``trustworthy'' and ``responsible'' AI, overlooks a more fundamental human-computer interaction (HCI) crisis: the erosion of human agency. 
This paper argues that the primary challenge of high-stakes AI systems is not trust, but the preservation of human causal control. 
We posit that ``bad AI'' will function as ``bad UI,'' a metaphor for catastrophic interface failures that misrepresent system state and lead to human error.
Applying Marshall McLuhan’s media theory, AI can be framed as a technology of ``augmentation'' that simultaneously ``amputates'' the user's direct perception of causality. 
This places the interface as the critical locus where a ``double uncertainty''—that of the human user and that of the probabilistic model—must be mediated. 
We critique current Explainable AI (XAI) for its correlational focus and failure to represent uncertainty. 
We conclude by proposing a rigorous, nested Causal-Agency Framework (CAF) that integrates causal models, uncertainty quantification, and human-centered evaluation to restore agency at the interface.
\end{abstract}




\maketitle

\section{Introduction}

Decades of human-computer interaction (HCI) and human factors research have demonstrated one critical principle that can be fatal: Bad user interfaces kill people. The Therac-25 radiation therapy accidents of 1985 provide a canonical example~\cite{leveson1993investigation}. A software race condition combined with a cryptic, unresponsive UI resulted in massive patient overdoses. Operators faced with an uninterpretable ``Malfunction 54'' error were desensitized by frequent minor bugs and simply proceeded to re-administer lethal doses.

This was not an isolated phenomenon. Interface failures have repeatedly resulted in catastrophic design errors across high-stakes domains. In aviation, for example, poorly designed ``glass cockpit'' interfaces often fail to provide pilots with awareness of the state of automation, contributing to numerous controlled flight into terrain (CFIT) accidents~\cite{norman1990problem}. The 2009 Air France Flight 447 crash is another example of this. Pilots were unable to diagnose a deep stall due to confusing and unreliable airspeed data, which led to the loss of 228 lives. Similarly, the 1979 Three Mile Island nuclear incident was fundamentally an HCI failure. Operators were overwhelmed by an ``alarm flood'' and misled by an ambiguous indicator light that incorrectly signaled that a critical valve had closed~\cite{rasmussen2012skills}. In a military context, the 1988 downing of Iran Air Flight 655 by the USS Vincennes highlights how the Aegis Combat System's complex interface ambiguously displayed the aircraft's altitude and identity. This led the crew to misinterpret the civilian airliner as a hostile target, resulting in a fatal misinterpretation. Furthermore, poor interfaces even jeopardize modern medical practice. Poorly designed electronic health record (EHR) systems lead to dangerous ``alert fatigue,'' contributing to medication and diagnostic errors~\cite{10.1001/jama.2018.1171}.

In all these cases, the failure was not due to ``human error'' alone; rather, it was caused by a design flaw, or the design itself hindered the recognition of an error. The interface created a fatal ambiguity between human intention and machine execution. As we embed complex, probabilistic AI models into critical infrastructure, we are not just scaling this problem—we are changing its nature. ``Bad AI'' is the new ``bad UI,'' but with exponentially greater catastrophic potential. The prevailing ethical discourse focused on ``responsible'' or ``trustworthy'' AI is a dangerous distraction. Rather than fostering blind trust, the goal should be to design for human agency—the user's ability to act as the ``causal agent'' of an event~\cite{10.5555/38407}. This requires the user to be able to form an intention, execute an action, and—critically—perceive and understand the outcome of that action.

The prevailing ethical discourse, focused on ``responsible'' or ``trustworthy'' AI, is a dangerous distraction. It frames the problem as one of believing in the system. This is a repeat of the Therac-25 error. Designing for user agency requires that the user can form an intention, execute an action, and—critically—perceive and understand the outcome of that action.


\section{The Crisis of Agency via Double Uncertainty}
As posited by~\citet{mcluhan1988understanding}, all technologies are extensions of the human body. The automobile can be regarded as an augmentation of the foot, and the telescope, as an augmentation of the eye. However, it is important to note that every augmentation is accompanied by an inherent loss, which can be considered an ``amputation.'' The automobile has effectively rendered the need for a walking culture obsolete, and the telephone has done the same for the art of penmanship~\cite{mcluhan1994understanding}.
AI can be regarded as the ultimate McLuhan-esque medium. It is an augmentation of the human nervous system and cognitive processes, a concept empirically supported by a growing body of research on cognitive offloading. Humans increasingly treat external digital tools as a form of transactive memory, thereby offloading the task of remembering.
This phenomenon, termed the ``Google Effect'' or ``digital amnesia,'' has been identified as a measurable tendency to forget information that is believed to be readily accessible online~\cite{doi:10.1126/science.1207745}.

However, it is imperative to delineate the specific aspects of cognitive function that are impacted by this augmentation. One can posit that this phenomenon is more fundamental than mere factual recall. This results in the loss of direct access to the causal chain by the user. The utilization of a rudimentary apparatus engenders a straightforward, corporeal awareness of the interplay between action and reaction. The utilization of a sophisticated AI system can result in the ``amputation'' of the user's intention from its consequential outcome. This process is mediated by a system characterized by intricacy in its internal logic, which is often opaque and non-deterministic in nature. As previously mentioned, all elements are ``played at the interface.'' The interface, in this case, becomes a phantom limb, a space where the user's sense of control and effect is no longer grounded in direct, verifiable perception.
This ``amputation'' of causality creates a crisis for human agency, defined by a ``double uncertainty'': a model uncertainty and a user uncertainty, respectively.

\begin{itemize}
\item Model Uncertainty: The AI, as a probabilistic system, is inherently uncertain. It does not ``know'' in a human sense; it calculates statistical likelihoods. This includes aleatoric uncertainty (noise in the data) and epistemic uncertainty (the model's own ignorance).
\item User Uncertainty: The human, confronted with this opaque system, is uncertain about what the AI is doing, why it is doing it, and what the effect of their own intervention will be.
\end{itemize}

This ``uncertainty behind an uncertain model'' paralyzes the feedback loop required for situated action~\cite{10.5555/38407}. 
The user cannot form a clear plan or interpret the system's output, preventing them from acting as a competent causal agent. 
This is where ``human-machine'' collaboration breaks down and design-induced error becomes inevitable, a phenomenon reflected in recent meta-analyses showing that human-AI combinations frequently fail to outperform AI alone~\cite{vaccaro2024combinations}.

\section{The Limits of Explainable Artificial Intelligence}
The field of explainable artificial intelligence (XAI) emerged as a technical and human-centered response to the opacity of complex models. The ``XAI Question Bank'' (XAIQB)~\cite{10.1145/3313831.3376590} represented a substantial advancement by transitioning from a primarily technical model-interrogation approach to a user-centric methodology, formalizing the types of inquiries that actual users pose (e.g.,~''Why did you do that?'', ``What if I did this?''). However, as the user's notes identified, and as subsequent research applying the XAIQB in practice has confirmed~\cite{10.1145/3603555.3608551}, significant gaps remain between the demand for explanations and the supply of truly useful, agentic tools.
Current XAI practices largely fail because they inadequately address the three pillars of human agency in high-stakes systems: causality, uncertainty, and actionability.

\textbf{Move from Correlational to Causal Explanations}: Most popular XAI methods (e.g.,~LIME, SHAP) are correlational. They identify which features were most important for a decision, but they cannot explain the underlying causal mechanism~\cite{pearl2018}. Answering ``Why?'' requires an understanding of cause and effect, not just feature attribution.
A truly agentic interface must move beyond the prevalent correlational paradigm. Most popular post-hoc XAI methods, such as LIME and SHAP, are essentially feature attribution techniques. These techniques identify which features in the data were most important for the model's decision-making process~\cite{rudin2019stop}. These methods answer the question: ``What features did the model correlate with this output?''

However, this is not what humans typically mean when they ask ``Why?'' Research in the social sciences and the philosophy of AI has shown that the human question ``Why?'' is almost always a causal and counterfactual inquiry~\cite{10.5555/3238230,10.1145/3313831.3376590}. The user is not asking, ``What features had high weights?'' Rather, they are asking, ``Why did this outcome happen instead of a different one?'' or ``What would I need to change to get a different outcome?''

Answering these questions requires an understanding of the underlying causal mechanism, not just the spurious correlations that the model may have learned~\cite{10.1007/978-3-031-04083-2_8}. A high SHAP value for ``fever'' in a diagnostic model only reveals the model's learned association. It cannot tell a user (e.g.,~a physician) whether the fever caused the condition or if both were symptoms of an unobserved third factor. In order to act as an agent, a human must be able to reason about interventions~\cite{10.5555/3238230,10.1007/978-3-031-04083-2_8}. Current XAI fails to provide the necessary tools, offering only correlational shadows.

\textbf{Represent and Propagate Uncertainty}: A second critical failure is that XAI explanations are often presented with a brittle, unearned confidence. They fail to communicate the model's own uncertainty, which can be of two types: the model's own uncertainty (epistemic) and the data's inherent limitations (aleatoric)~\cite{10.1007/978-3-031-04083-2_8,10.5555/3618408.3619058}. 
A user cannot make a risk-informed decision if the system presents a low-confidence ``guess'' as a high-confidence ``fact''~\cite{dubey2025ubiqtree}. 
In such cases, the XAI explanations themselves are unstable. 
In other words, the feature with the highest contribution may not be the most effective or reliably predictive.
In order to provide users with an accurate representation of the system's state, uncertainty must be propagated from the model through the explanation to the interface~\cite{10.1007/978-3-031-04083-2_8}.

\begin{table*}
\caption{Necessary Capabilities for Agency-Centric XAI: Moving Beyond Correlational Explanations to Support Human Intervention.}  \label{tab:xai_caps}
  \begin{tabular}{p{4.0cm} p{6.5cm} p{4.75cm}} 
    \toprule
    \textbf{Core Capability} & \textbf{Design Imperative} & \textbf{Key Supporting Literature} \\
    \midrule
    \textbf{Causal Reasoning} & Transition XAI outputs from correlation (feature attribution) to causal mechanisms (intervention and counterfactuals). &\citet{10.5555/3238230};~\citet{10.1007/978-3-031-04083-2_8};~\citet{10.1145/3313831.3376590}\\
    \textbf{Uncertainty Quantification} & Systematically communicate the model's epistemic (confidence) and aleatoric (data noise) risks to enable risk-informed user decisions. &\citet{10.1007/978-3-031-04083-2_8};~\citet{dubey2025ubiqtree} \\
    \textbf{Actionable Utility} & Shift evaluation focus from passive user comprehension and readability to active utility for user intervention and task goal achievement. &\citet{10.5555/38407};~\citet{10.1145/3603555.3608551};~\citet{ilgen-hattab-2025-toward} \\
    \bottomrule
  \end{tabular}
\end{table*}

\textbf{Evolve from Readability to Actionability}: The evaluation of XAI often stops at insufficient, low-level metrics like human comprehension or readability. Success is often defined as ``Does this explanation seem satisfying?'' or ``Can the user predict the model's output?''~\cite{10.1145/3313831.3376590}. As~\citet{ilgen-hattab-2025-toward} argue, this focus on ``readability'' is inadequate. It conflates the perceivability of an explanation with its cognitive utility and actionability. An explanation is only useful if it empowers the user to take a correct and effective action—to intervene, to correct, or to override. 
An explanation that is ``readable'' yet misleading is worse than no explanation at all. 
The metric for success is not ``Did the user understand?'' but ``Did the user+AI system achieve the goal safely?''
This aligns with the ``situated action'' paradigm~\cite{10.5555/38407}, in which the true test of an interface is its usefulness in real-world tasks. The objective is actionable agency, not user satisfaction (a proxy).

\section{The Causal-Agency Framework (CAF)}

To restore agency, we must build systems that are designed for causal and uncertainty-aware interactions and that require the three core capabilities presented in Table~\ref{tab:xai_caps}. 
We propose a Causal-Agency Framework (CAF), a ``nested model'' similarly to the nested model of AI design and validation but that handles these factors ``across the board''~\cite{dubey2024nested}. 
This framework consists of three integrated, co-designed layers:

\subsection{Causal \& Uncertainty Quantification (CUQ) Engine}

This is the technical backend. It must be built from the ground up to move beyond standard pattern recognition.
\begin{itemize}
\item Causal Modeling: The engine must utilize Structural Causal Models (SCMs) or causal discovery algorithms to approximate the underlying generative process~\cite{10.5555/3238230}. This allows the system to move from $P(Y|X)$ (correlation) to $P(Y|do(X))$ (intervention), enabling it to answer ``what if'' questions.

\item Uncertainty Quantification: The engine must rigorously quantify and propagate uncertainty at every step. 
Recent implementations such as UbiQTree~\cite{dubey2025ubiqtree} demonstrate the feasibility of this approach, effectively integrating uncertainty quantification into tree-based XAI to distinguish between aleatoric and epistemic risk.
Explicitly modeling both aleatoric and epistemic uncertainty makes the ``unknown unknowns'' visible to the layers above~\cite{10.5555/3618408.3619058}.
\end{itemize}

\begin{figure*}[t]
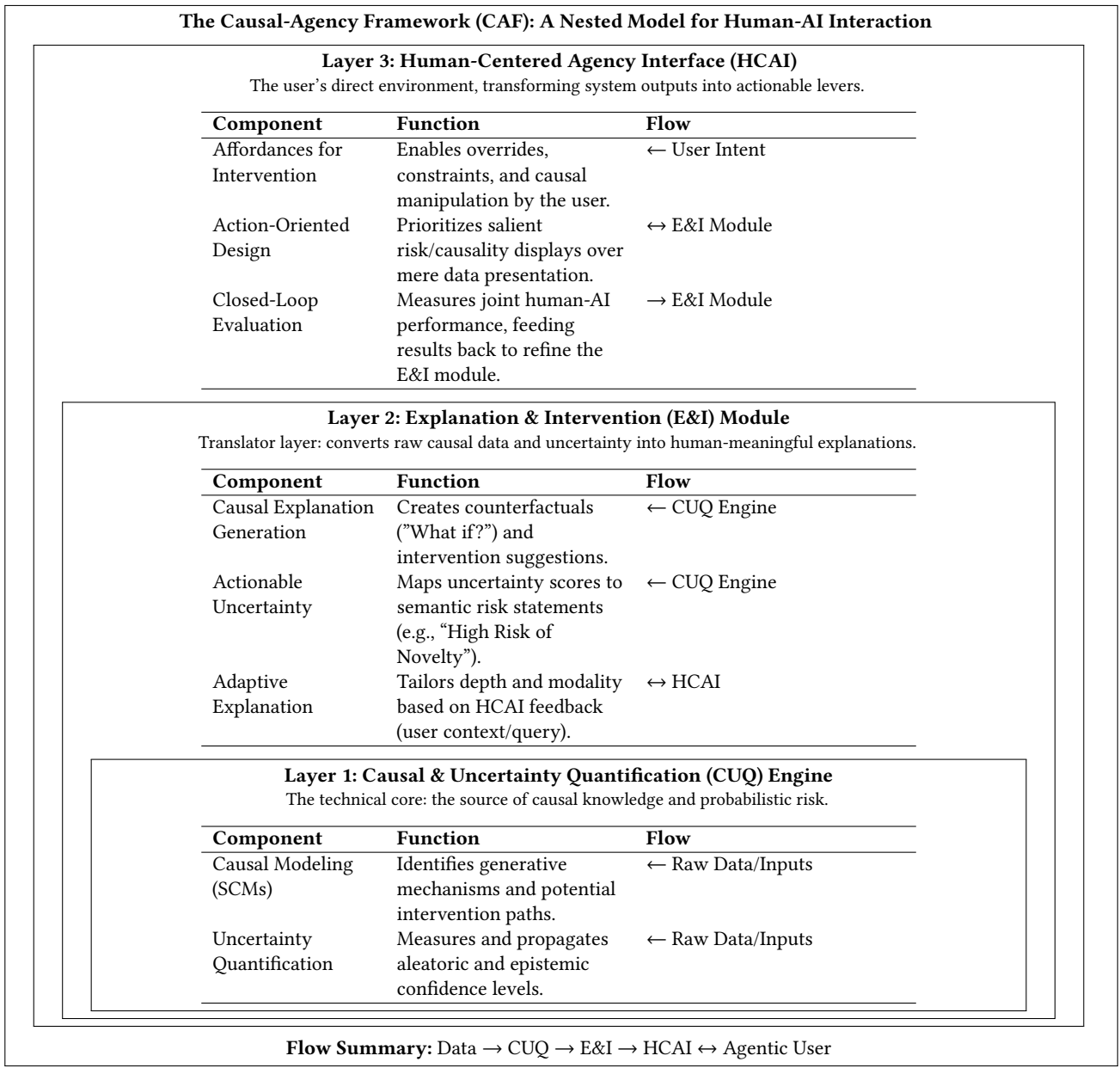

    \centering
    \fbox{\parbox{0.95\textwidth}{
    \centering
    \textbf{The Causal-Agency Framework (CAF): A Nested Model for Human-AI Interaction}
    \vspace{0.5em}
    
    \fbox{\parbox{0.9\textwidth}{
        \centering
        \textbf{Layer 3: Human-Centered Agency Interface (HCAI)}
        \vspace{0.2em}
        \parbox{0.8\textwidth}{\centering \small The user's direct environment, transforming system outputs into actionable levers.}
        \vspace{0.5em}
        
        \begin{tabular}{>{\raggedright\arraybackslash}p{2.5cm} >{\raggedright\arraybackslash}p{3.5cm} >{\raggedright\arraybackslash}p{4cm}}
            \hline
            \textbf{Component} & \textbf{Function} & \textbf{Flow} \\
            \hline
            Affordances for Intervention & Enables overrides, constraints, and causal manipulation by the user. & $\leftarrow$ User Intent \\
            Action-Oriented Design & Prioritizes salient risk/causality displays over mere data presentation. & $\leftrightarrow$ E\&I Module \\
            Closed-Loop Evaluation & Measures joint human-AI performance, feeding results back to refine the E\&I module. & $\rightarrow$ E\&I Module \\
            \hline
        \end{tabular}
        
        \vspace{0.5em}
        \fbox{\parbox{0.85\textwidth}{
            \centering
            \textbf{Layer 2: Explanation \& Intervention (E\&I) Module}
            \vspace{0.2em}
            \parbox{0.75\textwidth}{\centering \small Translator layer: converts raw causal data and uncertainty into human-meaningful explanations.}
            \vspace{0.5em}
            
            \begin{tabular}{>{\raggedright\arraybackslash}p{2.5cm} >{\raggedright\arraybackslash}p{3.5cm} >{\raggedright\arraybackslash}p{4cm}}
                \hline
                \textbf{Component} & \textbf{Function} & \textbf{Flow} \\
                \hline
                Causal Explanation Generation & Creates counterfactuals (''What if?'') and intervention suggestions. & $\leftarrow$ CUQ Engine \\
                Actionable Uncertainty & Maps uncertainty scores to semantic risk statements (e.g., ``High Risk of Novelty''). & $\leftarrow$ CUQ Engine \\
                Adaptive Explanation & Tailors depth and modality based on HCAI feedback (user context/query). & $\leftrightarrow$ HCAI \\
                \hline
            \end{tabular}
            
\vspace{0.5em}
            \fbox{\parbox{0.8\textwidth}{
                \centering
                \textbf{Layer 1: Causal \& Uncertainty Quantification (CUQ) Engine}
                \vspace{0.2em}
                \parbox{0.7\textwidth}{\centering \small The technical core: the source of causal knowledge and probabilistic risk.}
                \vspace{0.5em}
                
                \begin{tabular}{>{\raggedright\arraybackslash}p{2.5cm} >{\raggedright\arraybackslash}p{3.5cm} >{\raggedright\arraybackslash}p{4cm}}
                    \hline
                    \textbf{Component} & \textbf{Function} & \textbf{Flow} \\
                    \hline
                    Causal Modeling (SCMs) & Identifies generative mechanisms and potential intervention paths. & $\leftarrow$ Raw Data/Inputs \\
                    Uncertainty Quantification & Measures and propagates aleatoric and epistemic confidence levels. & $\leftarrow$ Raw Data/Inputs \\
                    \hline
                \end{tabular}
            }} 
            
        }} 
        
    }} 
    
    \vspace{0.5em}
    \centering
    \textbf{Flow Summary:} Data $\rightarrow$ CUQ $\rightarrow$ E\&I $\rightarrow$ HCAI $\leftrightarrow$ Agentic User
    
    } 
    }
    \caption{The Causal-Agency Framework (CAF). From the Data to the User: The flow moves from the core technical engine (CUQ) outwards through the translational module (E\&I) to the user's interface (HCAI), facilitating a closed-loop interaction with the Agentic User.}
\end{figure*}

\subsection{Explanation \& Intervention (E\&I) Module}

This layer acts as the system's ``translator'' and ``governor,'' mediating between the CUQ Engine and the user.
\begin{itemize}
\item Generates Causal Explanations: This module generates counterfactuals (''This outcome occurred because variable A was not B'') and causal explanations (''To change this outcome, you must intervene on A'').

\item Communicates Actionable Uncertainty: It translates raw uncertainty scores (e.g.,~variance) into semantic, actionable risk statements (''High confidence, data is typical,'' ``Low confidence, input is novel,'' or ``High data noise, recommend verification'').

\item Adaptive Explanation: The module must tailor the modality and complexity of the explanation to the user's cognitive state and the context's stakes, ensuring genuine comprehension and not just ``readability'' ({\.I}lgen \& Hattab, 2025).
\end{itemize}

\subsection{Human-Centered Agency Interface (HCAI)}

This is the user-facing layer where agency is ultimately won or lost. This interface is not a ``dashboard'' (read-only) but a ``cockpit'' (read-write).
\begin{itemize}
\item Affordances for Intervention: The HCAI must provide clear, unambiguous affordances for the user to act as the final causal agent. This includes the ability to override decisions, constrain the model's reasoning, demand new information, or provide new data (e.g.,~''ignore this feature,'' ``re-run with this hypothesis'').

\item Action-Oriented Design: The interface must be designed to facilitate action, not just trust. It must make the causal levers (from the E\&I module) and the associated risks (from the CUQ engine) salient and accessible.

\item Closed-Loop Evaluation: The success of the HCAI is not measured by user satisfaction or ``trust'' scores. It is measured by the performance, safety, and efficiency of the joint human-AI system in achieving its objective~\cite{dubey2024nested,10.5555/38407}.
\end{itemize}

\section{Discussion}
The proposed Causal-Agency Framework (CAF) does not constitute an incremental addition to existing XAI toolkits. This methodological argument calls for a fundamental re-architecture of high-stakes AI systems. It is grounded in a critique of current literature and practice. The implications of this phenomenon are examined through the lens of five primary axes.

Firstly, there is a shift from post-hoc explanation to \textit{a priori} design.
A substantial proportion of the extant literature on XAI, particularly in the context of practical applications, conceptualizes explainability as a post-hoc attribute applied to a pre-trained black-box model~\cite{10.1145/3313831.3376590}. This approach is inherently constrained; it endeavors to emulate the logic of an opaque system, which can yield explanations that are either unfaithful or misleading~\cite{rudin2019stop}. 
Even recent breakthroughs demonstrating that aligning human and machine abstractions can improve model generalization are insufficient for agency~\cite{muttenthaler2025aligning}. 
While ``aligned'' internal representations may enhance performance, they remain opaque to the user without an explicit interface to inspect and contest that alignment.
Conversely, the CAF employs the ``nested model'' structure~\cite{dubey2024nested}, which is an \textit{a priori} design paradigm. The system is interpretable by design at its core, due to the centralization of the Causal \& Uncertainty Quantification (CUQ) Engine. Causality and uncertainty are not merely afterthoughts; rather, they are fundamental computational principles. This methodological shift entails a transition from an approach that emphasizes ``trying to explain'' a system to one that prioritizes ``building a system that can explain itself'' through its causal architecture.

Secondly, it is imperative to distinguish between concerns. The technical, translational, and interactional layers are the fundamental components of this distinction.
A prevalent issue in the implementation of XAI is the conflation of explanation generation with its utility. The CAF methodologically separates these concerns into three distinct layers.
The CUQ Engine resolves the fundamental technical challenge of causal discovery and uncertainty quantification, which is currently a subject of active research~\cite{10.5555/3238230,10.5555/3618408.3619058}.
While historically difficult, emerging architectures like UbiQTree~\cite{dubey2025ubiqtree} provide the necessary computational foundation for this layer.
Nevertheless, a practical gap remains: current methodologies are inadequate for addressing causality across all levels and varying clinical contexts. Therefore, the realization of the CAF depends on developing newer, more robust causal methods that go beyond existing correlational protocols to support the engine's rigorous demands.
This separation enables interface designers to more effectively focus on communication during subsequent stages.
The E\&I Module addresses the translational problem by serving as a mediator, converting complex statistical and causal outputs into human-centric risk statements and counterfactuals.
The HCAI addresses the interaction problem by assessing the efficacy of the translation in restoring agency.
This separation is critical because it allows for specialized research and validation at each layer. We can advance causal discovery methods (Layer 1) while simultaneously running human-factors experiments on the most effective way to visualize uncertainty (Layer 3), without the two confounding each other.

Thirdly, it is imperative to adopt a novel approach to evaluation, shifting the focus from the concept of ``trust'' to that of ``joint system performance.''
A review of the extant literature reveals a plethora of ``responsible AI'' frameworks that prioritize the cultivation of ``trust.'' This is a flawed metric. The ``automation bias'' phenomenon—wherein humans unduly over-trust an automated system—is a well-documented source of error~\cite{doi:10.1518/001872097778543886}. A system that fosters unquestioning trust can be classified as a ``bad UI,'' according to our definition. The Therac-25 operator interpreted the ``Malfunction 54'' message as a benign, resettable error.
Recent research confirms that AI can improve user performance while simultaneously degrading their metacognitive understanding (User is ``smarter but none the wiser'')~\cite{FERNANDES2026108779}. This finding empirically validates the ``McLuhan amputation'' of causal awareness. Furthermore, this same research suggests that attempts to mitigate this disconnect through simple interaction framing (e.g.,~'AI as teammate' vs. 'AI as tool') are ineffective. This reinforces our central thesis: agency is not a psychological ``framing'' problem but a structural design challenge that demands the causal and uncertainty-aware architecture of frameworks such as the CAF.
The evaluation of the Human-Centered Agency Interface (HCAI) does not entail an assessment of user trust in the AI system; rather, it focuses on the efficacy and safety of the decisions made by the integrated human-AI system. This approach aligns with the ``situated action'' paradigm~\cite{10.5555/38407} and necessitates a rigorous, closed-loop evaluation of task performance~\cite{dubey2024nested}. The objective is not to develop a ``trustworthy'' system, but rather a system that is both scrutable and controllable, thereby enhancing the competencies of the human operator.

Fourthly, the shift to agency-centric interfaces (e.g.,~HCAI) and causal reasoning based frameworks (e.g.,~CAF) requires an overhaul of training protocols for domain experts. Although the existing literature emphasizes the importance of human expertise in validating AI recommendations and preventing inaccurate advice~\cite{doi:10.2105/AJPH.2024.307888}, implementing CAF requires that domain experts receive specialized training that goes beyond basic data, statistical, and digital literacy~\cite{Vermeire2025,gal2002adults,9975362,10.1108/LHTN-12-2021-0095}. This training must focus on the principles of causal inference and uncertainty interpretation, as presented by the E\&I module. Without instruction on how to use the HCAI's `Affordances for Intervention', which serve as the technical mechanisms for the `redress' central to human agency~\cite{fanni2023enhancing}, including how to manipulate causal levers and act on the quantification of epistemic risk, the system faces the dangers of misuse, rejection, and rendering the human expert incompetent. Thus, successful agency restoration is contingent upon building human competency to match the system’s transparency.

Fifth and finally, pursuing agency requires critically reevaluating data acquisition and evaluation strategies. One might question the feasibility of shifting to causal XAI using the purely observational ``Big Data'' currently employed in AI systems. The detection of hidden confounders, a prerequisite for valid causal inference, often requires interventional data or rich metadata, which is usually absent from standard training sets. Therefore, implementing the CAF implies a ``data imperative'': we must transition from indiscriminately collecting more data to strategically collecting ``causal'' data. Without validity at the data layer, even the most sophisticated agency-centric interface could present users with ``causal mirages'' rather than actionable realities.

\section{Conclusion}
The fatal flaws of the Therac-25, which masked lethal complexity behind an ambiguous interface, stand as a warning. As we deploy AI in critical domains, we are building interfaces for systems with greater complexity. Continuing to frame this challenge as one of ``trust'' is a moral and engineering error; it is a recipe for repeating past failures.
The inability to understand and direct causal effects of one's actions renders one a passenger, not an agent—and in high-stakes situations, it can lead to fatal mistakes.
We believe that humans should maintain control, and this work proposes a method to do so. The Causal-Agency Framework (CAF) does not rely on post-hoc explanations. Instead, it uses causality and uncertainty quantification. Success is evaluated using system performance. This will help us move beyond responsible AI and toward practical interfaces that restore human control in high stakes domains.


\begin{acks}
To Akshat, for the pizza and discussing the limitations of the uncertainty embedded in Shapley additive explanations.
\end{acks}

\bibliographystyle{ACM-Reference-Format}
\bibliography{sample-base}

\end{document}